# Risk Aware Optimization of Water Sensor Placement

Antonio Candelieri
University of Milano-Bicocca,
Department of Economics,
Management and Statistics
Italy
antonio.candelieri@unimib.it

Andrea Ponti
University of Milano-Bicocca,
Department of Computer Science,
Systems and Communication
Italy
a.ponti5@campus.unimib.it

Francesco Archetti
University of Milano-Bicocca,
Department of Computer Science,
Systems and Communication
Italy
francesco.archetti@unimib.it

## ABSTRACT

Optimal sensor placement (SP) usually minimizes an impact measure, such as the amount of contaminated water or the number of inhabitants affected before detection. The common choice is to minimize the minimum detection time (MDT) averaged over a set of contamination events, with contaminant injected at a different location. Given a SP, propagation is simulated through a hydraulic software model of the network to obtain spatio-temporal concentrations and the average MDT. Searching for an optimal SP is NP-hard: even for mid-size networks, efficient search methods are required, among which evolutionary approaches are often used. A bi-objective formalization is proposed: minimizing the average MDT and its standard deviation, that is the risk to detect some contamination event too late than the average MDT. We propose a data structure (sort of spatio-temporal heatmap) collecting simulation outcomes for every SP and particularly suitable for evolutionary optimization. Indeed, the proposed data structure enabled a convergence analysis of a population-based algorithm, leading to the identification of indicators for detecting problem-specific converge issues which could be generalized to other similar problems. We used Pymoo, a recent Python framework flexible enough to incorporate our problem specific termination criterion. Results on a benchmark and a real-world network are presented.

## CCS CONCEPTS

• **Computing methodologies** → Machine learning, Modeling and simulation; • **Applied computing** → Operations research.

## KEYWORDS

Sensor placement, water network, multi-objective optimization, evolutionary optimization.

## 1 INTRODUCTION

### 1.1 Motivation

In this paper we are concerned with the identification of optimal "sensing spots", within a network, for effectively and efficiently monitoring the spread of "effects" triggered by "events". Many real-world problems fit into this general framework. This paper specifically addresses drinking water distribution networks (WDNs), where sensing spots are sensors deployed at specific locations (i.e., nodes of the network), events are natural/intended contaminations and effects are spatio-temporal concentrations of the contaminant. This problem is usually known as optimal Sensor Placement (SP), where the optimization goal is to minimizes an impact measure, such as the amount of contaminated water or the number of inhabitants affected before detection.

Another problem fitting naturally in this framework is the detection of "fake-news" on the web: posts are published by bloggers, are hyperlinked to other bloggers' posts and contents, and then cascade through the web. Posts are time stamped so we can observe their propagation in the blogosphere: in this setting the placement problem is to identify a small set of blogs which catch as many cascades as early as possible.

Water quality monitoring entails the cost of deploying and operating the sensors, so that the problem arises of finding the optimal SP for different number of devices, and to estimate the cost/benefit trade-off. Beside the cost of sensors there is a wide range of impact measures driving the SP and can be regarded as objective functions:

- time until a contamination event is detected
- expected population affected by an intrusion
- expected amount of contaminated water consumed
- the likelihood of detection.

Evaluating the above objective functions is expensive because it requires to perform hydraulic software simulation for a large set of contamination events, each one associated to a different location where the contaminant is injected at. The first impact measure is typically adopted (the others are based on it). Simulating the contaminant propagation allows to compute the Minimum Detection Time (MDT) provided by a specific SP over all the simulated events. Finally, the SP minimizing the average of the MDT is searched for. However, searching for an optimal SP is NP-hard: even for mid-size networks, efficient search methods are required, among which evolutionary approaches are often used.

The main motivation of this study is that considering only the average of MDT could be misleading. We propose a bi-objective formalization of the optimal SP problem, where the second objective is the minimization of the standard deviation of the MDT, that is associated to risk of detecting some contamination event too late with respect to the average MDT. Every SP is represented as a binary vector, leading to a bi-objective mixed-integer optimization problem. We have used Pymoo, a Python library for population-based multi-objective optimization.

## 1.2 Contributions

- A novel data structure to model and analyse spatio-temporal concentrations of a contaminant within a WDN to include into an optimal SP framework. Indeed, the important role of this data structure is that it is the basis of the computation of every possible impact metrics associated to a given SP.
- A bi-objective optimization formalization of the optimal SP problem, aiming at minimizing the mean and standard deviation of the minimum detection time over a set of different simulated contamination events. This leads to risk-aware Pareto-efficient SPs and allows the decision maker to choose the preferred balance between the earliest possible detection and the risk of a significant impact on customers.
- Definition of a so-called "information space", based on the proposed data structure. Indeed, it allows to map SPs from the search space (where each SP is represented as a binary vector) into the information space, whose elements are instances of the proposed data structure.
- An investigation, within the information space, of the causes of possible problem-specific convergence inefficiencies arising in solving the proposed bi-objective SP optimization problem through population-based algorithms.
- Definition of two "distance-based" performance indicators: one in the search space and one in the information space, aimed to monitor the "homogeneity" – in the two spaces, separately – of population along generations. Monitoring these two indicators allows to detect possible convergence inefficiencies and also suggests the design of novel and problem-specific cross-over and mutation operators.

## 1.3 Related Works

In 1.1 we briefly introduced the issue of optimal SP; here we provide a more specific analysis of prior work. There has been a significant research activity about optimal SP since the challenge "Battle of water sensors networks" [1]. Recent surveys are [2][3]. Here we shall focus only on some contributions based on mixed integer programming and evolutionary solution strategies.

In [4] a branch & bound algorithm is proposed based on greedy heuristics and convex relaxation. Another approach [5] models the optimal placement as a mixed-integer problem, solved through a linear programming solver and an adaptive heuristic. In [6] SP is addressed as a bi-objective optimization problem, with the objectives being the number of sensors and the contamination risk. An entirely different approach [7] proposes a submodular objective function and a greedy algorithm guaranteed to find at least a constant fraction of the optimal score.

Genetic Algorithms (GA) have a long history in multi-objective optimization, with many successful applications. NSGA-II, whose basic architecture has been proposed in [8], has become a de-facto standard for multi-objective optimization. It has evolved through many contributions [9-11] into a comprehensive and flexible computational framework, namely Pymoo [12]. Among the many applications of NSGA-II, beyond those quoted above in the specific water network domain, we mention [13] which considers a multiply constrained bounded knapsack problem arising in the resource allocation in scheduling of the casting in a foundry which is known to be NP-hard. The paper addresses 2 complexities, the dimension of the search space and its discreteness through a customized recombination operator. A general introduction on evolutionary computing with a focus on robotics and nonstationary problem is [14].

## 1.4 Organization of the Paper

Sect. 2 describes the optimal SP problem. Sect. 3 presents the hydraulic simulation software and the problem specific data structures proposed to organize the outputs of the simulations. Sect. 4 briefly recall multi-objective optimization, Pareto analysis and hypervolume. Sect. 5 summarizes some relevant information about termination criteria and performance indicators of the population-based algorithm used. Sect. 6 presents the experiments and the relevant results with respect to a toy example and a real-world WDN.

## 2 PROBLEM FORMULATION

A WDN can be modelled as a graph $G = (V, P)$ where vertices in $V$ represent junctions, tanks, reservoirs or consumption points, and edges in $P$ represent pipes, pumps, and valves.

We assume a set of possible locations for placing sensors, that is $L \subseteq V$. Thus, a SP is a subset of sensor locations, with the subset's size less or equal to $p$ depending on the available budget. An SP is represented by a binary vector $s\{0,1\}^{|L|}$ whose components are $s_i = 1$ if a sensor is located at node $i$, $s_i = 0$ otherwise. Thus, an SP is given by the nonzero components of $s$.

Let $A$ denote the set of contamination events which must be detected by a certain $s$, and $d_{ai}$ the impact measure associated to a contamination event $a \in A \subseteq V$ detected by the $i$th sensor.

A probability distribution is placed over possible contamination events associated to the nodes. In the computations we assume – as usual in the literature – a uniform distribution, but in general discrete distributions are also possible.

The impact measure can be anyone of those presented in 1.1; in this paper we consider $d_{ai}$ as the MDT. For each event $a$ and sensor placement $s$ the MDT is defined as $MDT_a = \min_{i\,:\,s_i=1} d_{ai}$.

Let denote with (P) the minimization of the average MDT:

$$(P) \begin{cases} \min f_1(s) = \sum_{a \in A} \alpha_a \sum_{i=1,\dots,|L|} d_{ai} x_{ai} \\ s.t. \\ \sum_{i=1,\dots,|L|} s_i \leq p \\ s_i \in \{0,1\} \end{cases}$$

where:

- $\alpha_a$ is the probability for the contaminant to enter the network at node $a$.
- $d_{ai}$ is the time required for a sensor located at node $i$ to detect the contaminant introduced at node $a$.

- $x_{ai} = 1$ if $s_i = 1$, where $i$ is the first sensor detecting the contaminant injected at node $a$; 0 otherwise.

This problem is identical to the well-known p-median facility location problem. Another possible formulation is related to the set covering problem [15].

In our study we assume that all the events have the same chance of happening, that is $\alpha_a = 1/|A|$, therefore $f_1(s)$ is simply computed as the average of MDTs:

$$f_1(s) = \frac{1}{|A|}\sum_{a \in A} \hat{t}_a \quad (1)$$

with $\hat{t}_a$ the minimum time step at which concentration reaches or exceeds a given threshold $\tau$ for the event $a$.
Now, let us consider the standard deviation of the MDTs, that is the second objective function we want to minimize – subject to the same set of constraints in (P). Since we assume that all the events have the same chance of happening, we can simply write the second objectives as:

$$\min f_2(s) = \sqrt{\frac{1}{|A|}\sum_{a \in A}(\hat{t}_a - f_1(s))^2} \quad (2)$$

## 3 SIMULATION, NETWORK AND EVENT DATA DESCRIPTION

### 3.1 WNTR

The Water Network Tool for Resilience (WNTR) [6] is a Python package designed to simulate and analyse resilience of WDNs. WNTR is based on EPANET 2.0, which is a tool to simulate flowing of drinking water constituents within a WDN. For each sensor placement $s$ the two objectives $f_1(s)$ and $f_2(s)$ are computed by running a WNTR simulation.

In our study, each simulation has been performed for 24 hours, with a simulation step of 1 hour. Assuming $L = V$ and $A = V$ (i.e., the most computationally demanding problem configuration) the time required to run a simulation for the toy-example called Net1 (available with EPANET and WNTR, and whose associated graph is depicted in Figure 1) is 2 seconds. The simulation time scales linearly with the inverse of the simulation step.

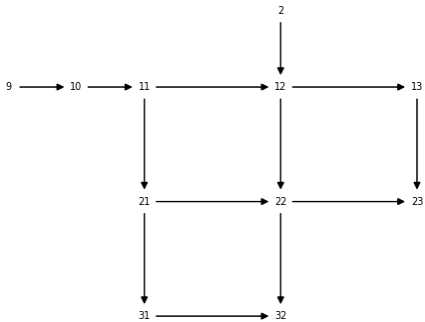

**Figure 1: A schematic representation of the Net1 toy-example**

### 3.2 Single Sensor and Sensor Placement Matrices

Let's denote with $S^\ell$ the so-called "sensor matrix", with $\ell = 1, ..., |L|$ an index identifying the location where the sensor is deployed at. Each entry $s^\ell_{ta}$ represents the concentration of the contaminant for the event $a \in A$ at the simulation step $t = 0, .., K$, with $T_{max} = K\Delta t$. Thus, in our study $T_{max} = 24$, $\Delta t = 1$ and $K = 24$. Without loss of generality, we assume that the contaminant is injected at the begin of the simulation (i.e., $t = 0$).

Analogously, a "sensor placement matrix", $H^{(s)} \in \mathbb{R}^{(K+1) \times |A|}$ is defined, where every entry $h_{ta}$ represents the maximum concentration over those detected by the sensors in $s$, for the event $a$ and at time step $t$. Suppose to have a sensor placement $s$ consisting of $m$ sensors with associated sensor matrices $S^1, ..., S^m$, then $h_{ta} = \max_{j=1,...,m} s^j_{ta} \ \forall a \in A$.

The two matrices can be depicted as heat maps, showing the different concentration dynamics, respectively for a single sensor and an SP. As example, Figure 2 shows the sensor matrices associated to two different sensors, while Figure 3 shows the sensor placement matrix resulting from the SP consisting of these two sensors.

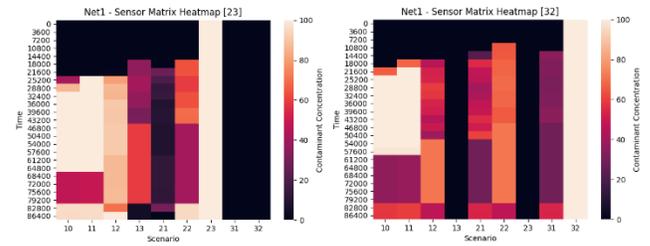

**Figure 2: Sensor matrices for sensors deployed at locations (i.e., nodes) 23 (left) and node 32 (right).**

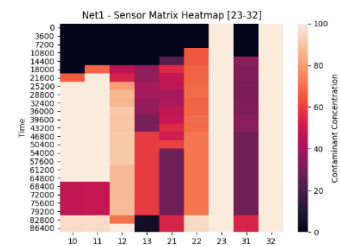

**Figure 3: Sensor placement matrix for a sensor placement consisting of two sensors deployed at locations 23 and 32.**

There is a relation between $s$ and the associated $H^{(s)}$: more precisely, the columns of $H^{(s)}$ having maximum concentration at row $t = 0$ (i.e., injection time) are those associated to events with injection occurring at the deployment locations of the sensors in $s$. Moreover, $H^{(s)}$ is the basic data structure on which MDT is computed. Indeed, we can now explicit the computation of $\hat{t}_a$ in the equations (1)(2) of $f_1(s)$ and $f_2(s)$: $\hat{t}_a$ is the minimum time step at which concentration reaches or exceeds a given threshold $\tau$ for the scenario $a$, that is $\hat{t}_a = \min_{t=1,\dots,K}\{h_{ta} \geq \tau\}$.

Unfortunately, the main issue is that we cannot work directly with sensor placement matrices. We are searching for an optimal vector $s$, while $H^{(s)}$ is just an additional observable information before computing the two objectives $f_1(s)$ and $f_2(s)$. This issue will be discussed in detail in the next section and it is at the core of the analysis presented in this paper.

# 4 Bi-objective optimization

## 4.1 Search space and information space

Our search space consists of all the possible SPs, given a set $L$ of possible locations for their deployment, and resulting feasible with respect to the constraints in (P). Formally, $s \in \Omega \subseteq \{0,1\}^{|L|}$. As already mentioned, the computation of the two objectives $f_1(s)$ and $f_2(s)$ requires the generation of the associated sensor placement matrix $H^{(s)}$. For the sake of simplicity, let's denote with $\pi$ this computational process:

$$s \xrightarrow{\pi} H^{(s)} \implies \phi(H^{(s)}) = (f_1(s), f_2(s))$$

We use $\phi(H^{(s)})$ to stress the fact that the computation is actually performed over $H^{(s)}$ – within the "information space" – and then it generates the observation of the two objectives $(f_1(s), f_2(s))$.

Any distance in $\Omega$ can be highly misleading, in that two SPs distant in $\Omega$ might correspond to a similar sensor placement matrix, leading to very close objectives values. This means that the landscape of the problem may have a huge number of global (not only local) optima, also significantly distant among them in $\Omega$. Indeed, supposed to have $s, s': d(x, x') = d_{max}$ (e.g., if $d(.,.)$ is the Hamming distance, $s = (0,1,0,1,\dots)$ e $s' = (1,0,1,0,\dots)$), then we could anyway observe $(f_1(s), f_2(s)) \cong (f_1(s'), f_2(s'))$ if $\delta(H, H') \cong 0$, with $H = H^{(s)}$, $H' = H^{(s')}$ and $\delta(.,.)$ a suitable distance between matrices. In this paper $\delta(H, H')$ is the Frobenius Norm of the matrix $H - H'$:

$$\delta(H, H') = \sqrt{\sum_{t=1}^{K}\sum_{a \in A}(h_{ti} - h'_{ti})^2}$$

A graphical representation of the specificities arising in this study is given in Figure 4.

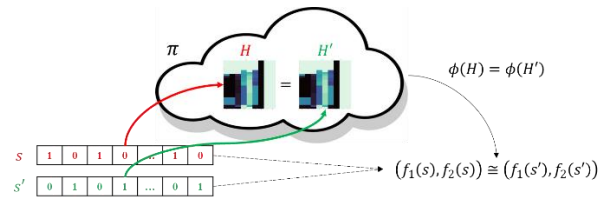

**Figure 4: Search space, information space and objectives**

The convergence in terms of objectives does not translate into a convergence in terms of the homogeneity of the population and, consequently, into a convergence to the optimum $s^*$. We will empirically prove this issue on a toy example for which the entire search space can be evaluated through exhaustive search.

## 4.2 Pareto analysis and the Hypervolume

Pareto rationality is the theoretical framework to analyse multi-objective optimization problems where $q$ objective functions $f_1(x), \dots, f_q(x)$, where $f_i(x): \to \mathbb{R}$ are to be simultaneously optimized in the search space $\Omega \subseteq \mathbb{R}^d$. Here we use $x$ to be compliant with the typical Pareto analysis's notation, clearly in this study $x$ is a sensor placement $s$.

We use the notation $\mathbf{f}(x) = \big(f_1(x), \dots, f_q(x)\big)$ to refer to the vector of all objectives evaluated at a location $x$. The goal in multi-objective optimization is to identify the Pareto frontier of $\mathbf{f}(x)$. To do this, an ordering relation in $\mathbb{R}^q$ is required, such that $\mathbf{f} = \big(f_1(x), \dots, f_q(x)\big) \preccurlyeq \mathbf{f}' = \big(f_1(x'), \dots, f_q(x')\big)$ if and only if $f_i(x) \leq f_i(x')$ for $i = 1, \dots, q$. This ordering relation induces an order in $\Omega: x \preccurlyeq x'$ if and only if $\mathbf{f}(x) \preccurlyeq \mathbf{f}(x')$.

We also say that $f(x')$ dominates $f(x)$ (strongly if $\exists\, i = 1, \dots, q$ for which $f_i(x) < f_i(x')$). The optimal non-dominated (aka *dominant*) solutions lay on the so-called Pareto frontier. The interest in finding locations $x$ having the associated $\mathbf{f}(x)$ on the Pareto frontier is clear: all of them represent efficient trade-offs between conflicting objectives and are the only ones, according to the Pareto rationality, to be considered.

A fundamental difference between single and multi-objective optimization is that it is not obvious which metric to use to evaluate the solution quality. A natural solution uses the so called hypervolume to compute the volume enclosed by a reference point dominated and the actual Pareto frontier. An example of Pareto frontier, along with the reference point to compute the hypervolume, is reported in Figure 5.

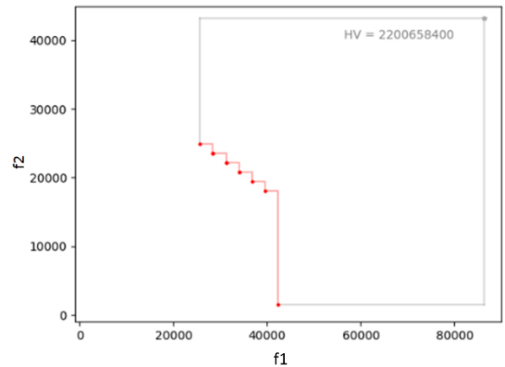

**Figure 5: An example of Pareto frontier, with the associated hypervolume, for two minimization objectives.**

Indeed, in multi-objective we aim at identifying the Pareto set, instead of a single optimal solution. A marginally/largely dominant

Pareto set will result into a low/high hypervolume value; thus, hypervolume is a reasonable measure for evaluating the quality of the optimization process.

## 5 OPTIMIZATION ALGORITHMS

### 5.1 Optimization Framework

The optimization framework used in this paper is Pymoo (Python multi-objective optimization) [12] based on the well-known NSGA-II algorithm [8]. Pymoo includes later versions of NSGA as well as other approaches. An important feature of Pymoo is that evolutionary operators and termination criteria are customizable.

### 5.2 Termination Criterion

Among others, a recent survey about termination criteria in evolutionary approach is given in [17]. Common termination criteria are a predetermined number of function evaluations, no-improvement for a prefixed number of consecutive generations, variability of the objective values in the population. Another relevant termination criterion proposed in the literature is the so-called Kappa criterion [17,18], which is fulfilled if the quotient of sum of all normalized distances between all individuals of the current population and $k_{max} = \frac{\mu^2-\mu}{2}$ is equal to or less than a given $\varepsilon > 0$, with $\mu$ the population size.

Our hypothesis is that Kappa values computed on the individuals – that are SP vectors $s$ – are larger than Kappa values computed on the associated sensor placement matrices $H^{(s)}$.

### 5.3 Performance Indicators

Pymoo provides two classes of performance indicators to monitor the progress of the optimization process, with respect the generation of dominant solutions. The two classes are: "distance-based", which evaluates the distance of the population from the Pareto frontier, and a "hypervolume-based".

In this study we extend the possible indicators by considering two further measures: Kappa computed in the search space and in the information space, respectively.

## 6 COMPUTATION RESULTS

### 6.1 Net1 toy-example

The toy-example named Net1 is a small WDN model provided by EPANET and WNTR, whose associated graph is depicted in previous Figure 1. Net1 consists of 1 reservoir (at location 9), 1 tank (at location 2) and 11 junctions (nodes). The set of possible locations to deploy sensors are the 11 junctions, therefore the set $L$ is $L = \{10, 11, 12, 13, 21, 22, 23, 31, 32\}$. The 11 junctions are also assumed to be the possible locations where the contaminant can be injected at, therefore $A = L$.

We have considered the case that the available budget allows to have a maximum of 4 sensors in the optimal SP, that is $p = 4$ in the constraints in (P) – valid also for the proposed bi-objective formulation. The value of $p$ has been defined according to a preliminary analysis on the single-objective problem (P): further increasing $p$ does not offer any further improvement of $f_1(s)$.

We remind that an SP is represented through a binary vector $s$ with $|L|$ components, with at the most $p$ components equal to 1. The search space for this toy example is therefore quite limited, allowing us to solve the problem via exhaustive search. More precisely, only 561 SPs are feasible according to the constraints in (P). Thus, we exactly know the Pareto set, the Pareto frontier and the associated hypervolume.

We have used Pymoo by setting 100 generations and a population size of 40 (for all the others NSGA-II's settings we have used the default values). Figure 6 shows the resulting Pareto frontier at the end of the optimization process, compared to the optimal Pareto (computed through exhaustive search).

Although the Pareto frontier identified by NSGA-II is very close to the optimal one (i.e., only one Pareto solution is missing), this experiment brings into light an issue due to ignoring which are the modifications in the information space implied by cross-over and mutation operations (which work in the search space). Indeed, the actual Pareto set consists of 36 out of 561 feasible SPs, but NSGA-II was able to identify only 8 of them, after 40 generations. Moreover, many solutions generated via NSGA-II are infeasible, and the only feasible solutions are the 8 Pareto solutions identified, that is just the 20% of the population size.

Figure 7 shows how the hypervolume (on the left) and the number of individuals whose associated objectives belong to the Pareto frontier (on the right) change along the NSGA-II generations.

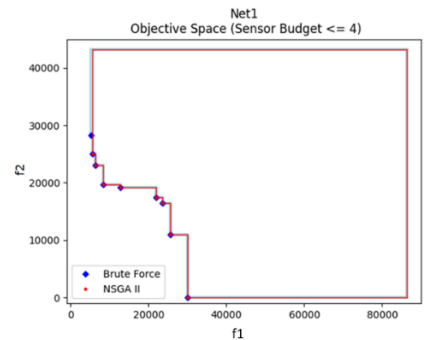

**Figure 6: Comparing approximated Pareto frontier by NSGA-II against the actual one computed through exhaustive search**

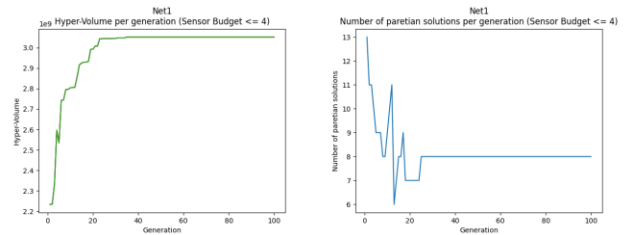

**Figure 7: Hypervolume (left) and number of Pareto set solutions within the population (right) over generations**

It is important to remark that, since the final user would like to choose, among the Pareto solutions and according to his/her preferences, the best trade-off of the two objectives, missing one Pareto solution (out of 9) means that we are anyway limiting his/her possible choices.

According to the identified issue, we have analysed the relations among search space, information space, values of the objectives and hypervolume. We have considered all the possible pairs $(s, s')$ among the 561 feasible SPs and computed the Hamming distances in the search space and the Frobenius distances in the information space. Figure 8 shows that there is a non-linear relation between Frobenius and Hamming (i.e., with Hamming integer-valued by definition). Furthermore, the standard deviation of the Frobenius distance decreases with the Hamming distance increasing, leading to have $(s, s')$ which are close in the search space (e.g., Hamming distance equal to 1) but with a large Frobenius distance, like that of an SP pair which is distant in the search space (e.g., Hamming distance equal to 8).

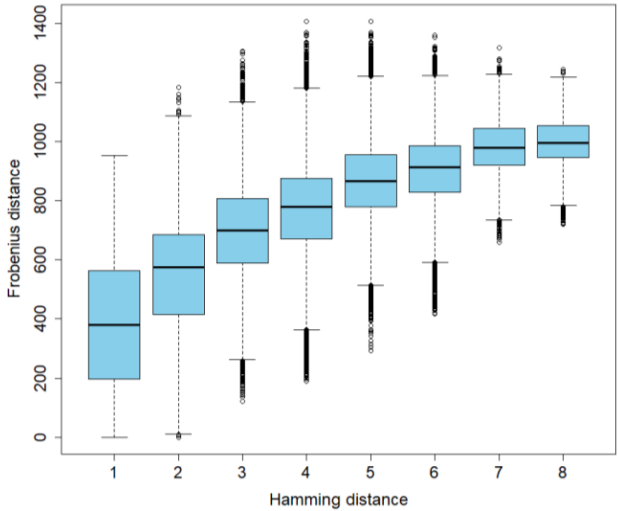

**Figure 8: Relation between search space (Hamming distance) and information space (Frobenius distance)**

We have then analysed the relationship between the distances in the two spaces and the two objectives, separately. Figure 9 shows that there is not any relevant variation in terms of $|f_1(s) - f_1(s')|$ depending on how much $s$ and $s'$ are distant in the search space (Hamming distance, on the left). Conversely – and as expected – SPs which are close (distant) in the information space have similar (different) values of $f_1$. Here we have discretized the Frobenius distance (observed in [0;1406.21]) into 8 equally sized bins, just to have a representation comparable with the Hamming's one.

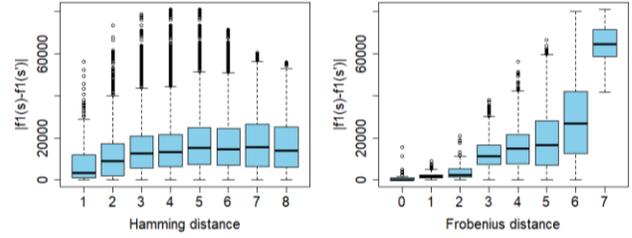

**Figure 9: Relation between $|f_1(s) - f_1(s')|$ and (left) Hamming distance and (righ) Frobenius distance**

Figure 10 shows similar results also for the second objective, even if the relation between Frobenius and $|f_1(s) - f_1(s')|$ is less evident.

All these results allow us to understand another important fact: although the optimal Pareto set (identified via exhaustive search) consists of 36 SPs, the associated Pareto frontier consists of only 9 points. Therefore, there is a relation many-to-one between the search space and the space spanned by the two objectives. These considerations can explain why Pymoo was not able – and cannot – identify the optimal Pareto frontier after 100 generations of 40 individuals each.

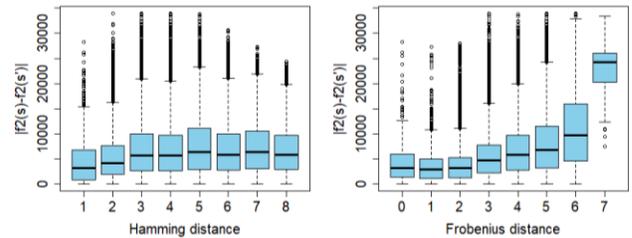

**Figure 10: Relation between $|f_2(s) - f_2(s')|$ and (left) Hamming distance and (right) Frobenius distance**

Finally, we have observed that min-max range of $f_1$ observed in the population along the generations (i.e., approximately 10'000-45'000), is incongruent with the actual Pareto frontier. This confirm that cross-over and mutation, *occurring in the search space*, do not allow to generate new dominant solutions, because *dominance is given by position within the information space*. The incongruence resulted less evident for the min-max range of $f_2$.

## 6.2   Neptun WDN

Neptun is small WDN in Timisoara, Romania, more specifically it is a District Metered Area (DMA) of a large WDN, and it was a pilot area of the European project ICeWater [19]. The graph associated to Neptun consists of 333 nodes and 339 edges (not reported here for space limitations).

We set $|L| = |A| = 332$ (i.e., we have excluded the graph node associated to the reservoir) and assumed $p = 25$ as the maximum number of sensors in a placement.

In this case study we cannot perform exhaustive search due to the huge number of possible SPs. Thus, we analysed the following set of indicators to understand which one of them could be used to possibly detect the issue investigated on the toy example:
- Hypervolume along generations
- Number of individuals belonging to the approximated Pareto frontier (we do not know the actual one) along generations
- Min-max range of $f_1$ along generations
- Min-max range of $f_2$ along generations
- Kappa with respect to the Hamming distance along generations
- Kappa with respect to the Frobenius distance along generations

The NSGA-II settings for this experiment are: population size equal to 40 and 500 generations (for all the other NSGA-II's setting we have used the default values).

Figure 11 shows the hypervolume of the approximated Pareto frontier for each generation. Up to approximately 200 generations, the population consists of infeasible solutions only, therefore hypervolume cannot be computed. Successively, hypervolume increases and converges in approximately 100 generations.

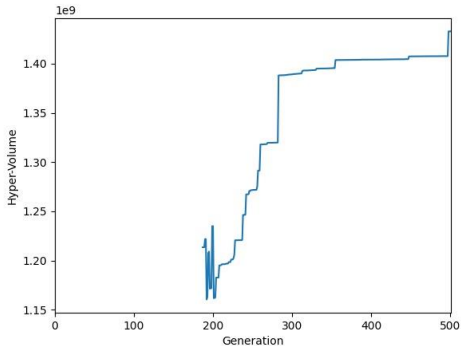

**Figure 11: Hypervolume for the approximated Pareto frontier at each generation**

Figure 12 displays, for each generation, the number of individuals that are feasible and whose values of $f_1$ and $f_2$ lay on the Pareto frontier. After 350 generations, approximately, this number does not significantly change any longer. According to results on the toy example, we cannot be sure that this is the number of points of the actual Pareto frontier.

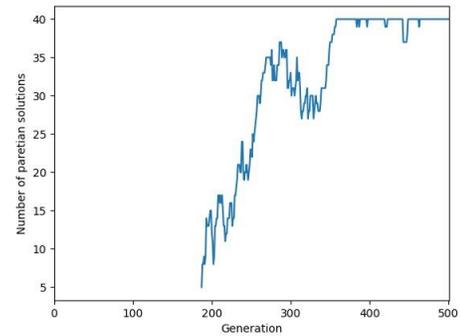

**Figure 12: Number of individuals in the population belonging to the approximated Pareto frontier**

Figure 13 displays the min-max range for the values of $f_1$ at each generation, computed for feasible and infeasible individuals, separately. Both the minimum and the maximum observed value of $f_1$ initially increases because the population consists of infeasible individuals, only. Starting from approximately 200 generations, feasible solutions appear in the population and the min-max range becomes wider (i.e., minimum starts to decrease).

In a specular way, Figure 14 shows that both the minimum and the maximum of $f_2$ decreases up to feasible individuals appear in the population, then the min-max range starts to increase (i.e., the maximum starts to increase).

These results assure us that we can finally provide a set of good decisions for our optimal SP problem, with around 40 different possible trade-offs (Figure 12) between $f_1$ and $f_2$.

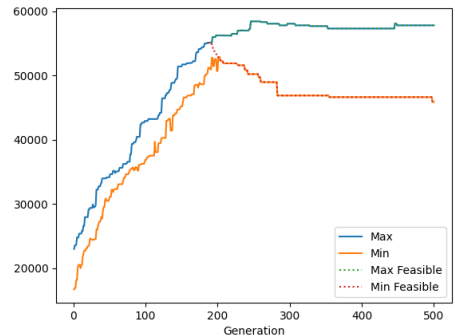

**Figure 13: Min-max range of $f_1$ at each generation**

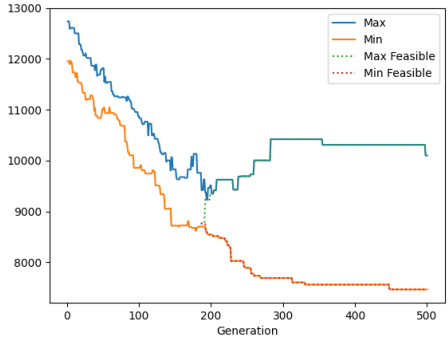

**Figure 14: Min-max range of $f_2$ at each generation**

However, we cannot be sure that the approximated Pareto frontier at the end of the 500 generations is a good approximation of the optimal one. If this approximation is poor then (a) there could be other solutions dominating our approximated Pareto frontier (i.e., hypervolume could be further increased) and (b) we could miss to provide the final user with trade-offs which he/she could prefer.

Our proposition is to investigate the Kappa indicators reported in Figure 15: a Kappa indicator computed into the search space (based on the Hamming distance, namely "Kappa-Hamming") and another Kappa indicator computed into the information space (based on the Frobenius, namely "Kappa-Frobenius").

While the population becomes more homogeneous within the search space (i.e., Kappa-Hamming decreases along generations), this is not true in the information space. Indeed, Kappa-Frobenius does not change significantly along generations, and it is quite larger than Kappa-Hamming – we want to remark that Kappa is a normalized indicator ranging in [0,1] therefore Kappa-Hamming and Kappa-Frobenius can be compared, even if the base distances vary in two different ranges.

Finally, we can conclude that also this real-life case study is affected by the issue investigated on the toy example.

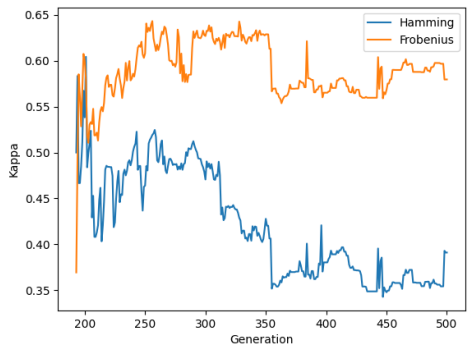

**Figure 15: Kappa indicator at each generation computed into the search space (Hamming distance) and information space (Frobenius distance), respectively**

As main result, we can conclude that the comparison between the two Kappa indicators – Kappa-Hamming and Kappa-Frobenius – is a reliable performance metric of the optimization process and, more importantly, a measure to detect a lack of convergence in the information space where the objectives are computed.

We are of the idea that a simple termination criterion on the Kappa-Frobenius is not sufficient to overcome the investigated issue. A more methodologically sound approach requires to define problem-specific cross-over and mutation operators to model, and exploit, the relationship between modifications carried onto the individuals (i.e., in the search space) and the corresponding "movements" in the information space.

## 7 CONCLUSIONS

Computational results offer evidence that not only suitable values for population size and number of generations are required to solve the proposed bi-objective optimal Sensor Placement (SP) problem, but also problem-specific indicators to detect convergence inefficiencies – to address through novel and problem-specific cross-over and mutation operators.

Analysis was enabled by a new data structure (sort of heatmap) modelling contamination dynamics within a Water Distribution Network (WDN) given a SP and a set of contamination events. This data structure revealed an information space – where objectives are computed – which has been investigated to understand possible causes of convergence inefficiencies.

Future work will address the definition of novel and problem-specific cross-over and mutation operators which can effectively consider information space to increase sample efficiency of NSGA-II.

It is important to remark that the proposed approach is naturally applicable to other problems sharing a network structure and/or spatio-temporal dynamics. For instance, in blogosphere monitoring one can select a set of blogs to promptly detect outbreaks, such as "fake news". Posts have rich metadata, including time stamps which allow to extract information about cascades (spatio-temporal propagation) and compute the Minimum Detection Time (MDT) of outbreaks. Other monitoring metrics fit in this framework as the detection likelihood in which the quality of the solution is the fraction of detected cascades and the population affected which depends on detection times but also on the size of cascades. These additional metrics could be incorporated in the optimal Sensor Placement problem, either as independent or multiple objectives. For instance, this approach could leverage tools, recently introduced by Facebook, that enable users to flag fake news, into the minimization of the spread of misinformation by detecting fake news in the network and stop their propagation.

Another application domain for the proposed approach could be monitoring urban vehicular traffic through the selection of the best subset of privately owned sensors to monitor the dynamics of vehicular traffic. In this case data refers to sensor readings and simulation models.